\documentclass[10pt]{spie}

\usepackage{algorithm,algorithmic,listings}
\usepackage{enumerate}
\usepackage{color,xcolor}
\usepackage{float}
\usepackage{amsmath}
\usepackage{graphicx}

\newcommand{\ket}[1]{\left| #1 \right>} % for Dirac bras
\definecolor{codebg}{HTML}{EEEEEE} % Background color used for code listings
\definecolor{figbdr}{HTML}{CCCCCC} % Used once for a figure border
\title{Programmable Multi-Node Quantum Network Design and Simulation}
\author{Venkat R.~Dasari\supit{a}, Ronald J.~Sadlier\supit{b}, Ryan Prout\supit{b}, \\ Brian P.~Williams\supit{b}, and Travis S.~Humble\supit{b}
\skiplinehalf
\supit{a}Army Research Laboratory, Aberdeen Proving Ground, Maryland; \\
\supit{b}Quantum Computing Institute, Oak Ridge National Laboratory, Oak Ridge, Tennessee
}

\authorinfo{This manuscript has been authored by the U.S Army Research Laboratory and UT-Battelle, LLC, under Contract No. DE-AC0500OR22725 with the U.S. Department of Energy. The United States Government retains and the publisher, by accepting the article for publication, acknowledges that the United States Government retains a non-exclusive, paid-up, irrevocable, world-wide license to publish or reproduce the published form of this manuscript, or allow others to do so, for the United States Government purposes. The Department of Energy will provide public access to these results of federally sponsored research in accordance with the DOE Public Access Plan. Further author information: \\ V.R.D.: E-mail: venkateswara.r.dasari.civ@mail.mil, Telephone: 1 410 278 2846 \\T.S.H.: E-mail: humblets@ornl.gov, Telephone: 1 865 574 6162}

\begin{document}
\maketitle

\begin{abstract}
Software-defined networking offers a device-agnostic programmable framework to encode new network functions. Externally centralized control plane intelligence allows programmers to write network applications and to build functional network designs. OpenFlow is a key protocol widely adopted to build programmable networks because of its programmability, flexibility and ability to interconnect heterogeneous network devices. We simulate the functional topology of a multi-node quantum network that uses programmable network principles to manage quantum metadata for protocols such as teleportation, superdense coding, and quantum key distribution. We first show how the OpenFlow protocol can manage the quantum metadata needed to control the quantum channel. We then use numerical simulation to demonstrate robust programmability of a quantum switch via the OpenFlow network controller while executing an application of superdense coding. We describe the software framework implemented to carry out these simulations and we discuss near-term efforts to realize these applications.
\end{abstract}


\section{Introduction}
A quantum network is composed of interacting nodes that express applications in quantum communication, computation and sensing \cite{VanMeter2014}. Its nodes and links represent the fundamental structure of the network with nodes containing both quantum and classical resources while links support both quantum and classical communication. Each device within the network has a role in the encoding, decoding, transmitting, receiving, repeating and routing of quantum information. Nodes may be further specialized to accommodate other specific tasks. For example, quantum computers may be viewed as quantum networks over short length scales specialized to perform computational tasks \cite{Britt2015}. Other specific applications that use quantum networks include quantum key distribution for secure communication \cite{Gisin2002}, blind quantum computing for secure interactive computation \cite{Broadbent2009}, and distributed quantum sensing for infrastructure protection \cite{Humble2013,Williams2015}. The advantages offered by quantum networks for solving these problems include greater utility, greater extensibility, and greater application resiliency.
\par
Despite the progress made in deploying specialized quantum networks, a desired feature for future quantum networks is the ability to easily encode a device-agnostic unified control plane capable of brokering communications between various quantum network node types. The advantage of programmable networks is that they can be tasked to support new functionality, for example, by converting between communication and sensing applications. In general, programmable networks reduce the overhead for future network deployment and permit rapid adoption of new functional paradigms \cite{Campbell1999}. Within the conventional networking community, programmability is a key advantage afforded by software-defined networking (SDN) principles \cite{Hu}. The SDN paradigm supports nodes within a network that can be managed by a external controller via software interfaces as opposed to pre-configured hardware constructions. One of the primary advantages of this programmable approach is that the nodes within a network can be programmed during deployment instead of resorting to costly hardware redesign or replacement. 
\par
Previously, we have shown how the basic principles of programmable networks can be extended to quantum networks, in which quantum metadata specifies attributes needed by the network to accommodate specific uses of the quantum channel \cite{HumbleSadlier2014,Dasari2015}. Quantum metadata is the classical information that moves through the network to characterize how applications make use of quantum data. This represents an initial step toward the goal of network programmability. Network programmablilty is realized by exchanging metadata about the requested functionality between a network controller and the network nodes, especially the switches and routers responsible for managing traffic movement. The movement of metadata is managed by the SDN controller, which implements policies for different metadata instances. Each type of network device may have different metadata requirements, e.g., a router versus a transmitter \cite{Dasari2015}. The SDN controller is responsible for distributing the metadata needed by each device to achieve a given functionality. This requires programmable devices within the quantum network. 
\par
In this contribution, we provide the first demonstration of a programmable quantum network by realizing the necessary interfaces for programmable quantum nodes and links. We clarify the role of quantum metadata within the programmable network paradigm by showing compatibility with the OpenFlow protocol, a widely used SDN paradigm for network control. We investigate how quantum networks can be built as programmable networks and how the OpenFlow protocol can be extended to account for metadata specific to quantum networks. Because of its programmability and compatibility with management of optical networks, OpenFlow is highly suitable to control the new attributes defining the quantum channels that carry metadata between various quantum devices. We build on this compatibility by designing models of the devices, including hosts and switches, needed to implement a quantum network. We then use numerical simulation alongside explicit classical networking to mimic the behavior of a switched quantum network performing super-dense coding. We implement quantum-classical communication protocols, including metadata exchanges, and we use numerical simulations that include noisy quantum transmissions. We present details of the quantum network simulator as well as the software infrastructure that can be easily extended for use in experimental test beds.
%%%%%%%%%%%%%%%%%%%%%%%%%%%%%%%%%%%
\begin{figure}[t]
\centering
\includegraphics[width=0.8\columnwidth]{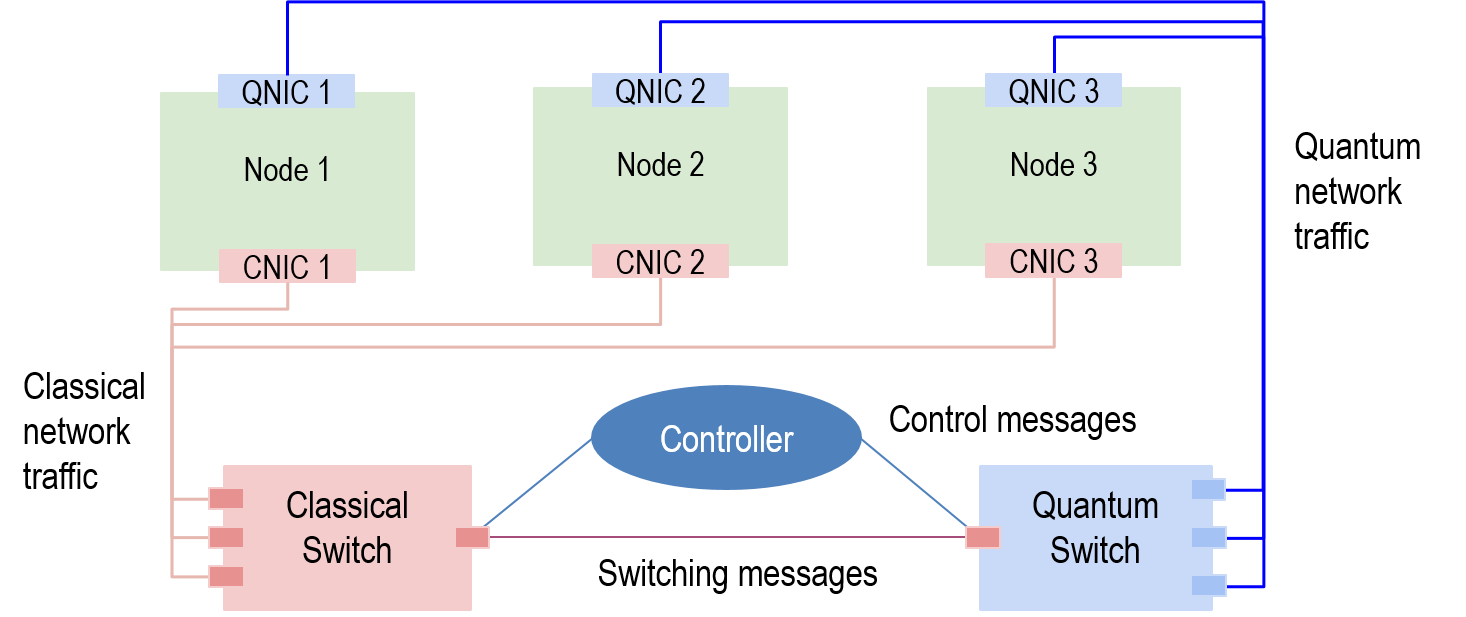}\\
\caption{Design schematic for a three-node quantum network with SDN controller connected to both the classical and quantum switch.}
\label{fig:quantumnetworkdesign}
\end{figure}
%%%%%%%%%%%%%%%%%%%%%%%%%%%%%%%%%%%
\section{Quantum Network Model}
A schematic design of a 3-node quantum network is shown in Fig.~\ref{fig:quantumnetworkdesign}. This diagram consists of three host nodes that each have a classical network interface card (CNIC) and a quantum network interface card (QNIC). These network interfaces connect a host to classical and quantum links that terminate at classical and quantum switches, respectively. We distinguish between classical and quantum network traffic as a natural separation of concerns in the function of a quantum network. Whereas the quantum network layer is responsible for transmitting quantum states between devices, the classical network layer is responsible for controlling those quantum transmission by passing metadata between the nodes. 
\par
As shown in Fig.~\ref{fig:quantumnetworkdesign}, the classical and quantum switches are linked directly with each other as well as with the Openflow controller. In particular, the quantum switch must have a CNIC to accept classical metadata from the hosts and to accept programming from the network controller. The controller manages interactions between the classical and quantum network layers by defining how and which metadata within the classical network should be forwarded to the quantum switch. For a quantum switch that supports teleportation or entanglement swapping, i.e., a quantum repeater, the controller may also perform functions that program the teleportation protocol. 
In general, the QNIC's and the quantum switch exchange quantum network traffic, which implies the structured transmission of quantum information carriers. Our implementation focuses on transmissions constructed from quantum optical carriers and especially from single and bi-photon states. The classical network traffic represents information encoded and transmitted using conventional network standards, e.g., TCP.
\par
The role of the SDN controller is highlighted in Fig.~\ref{fig:qnode}, which shows the interaction between a quantum network device and the OpenFlow controller used in our implementation. An agent within each device listens for requests from the controller that come in the form of periodic polling to monitor changes in device state. In addition, device policies may be set so that updates are pushed asynchronously to the controller. Higher-level networking functions, such as packetization, are set by the local device configuration, but the OpenFlow agents provides an interface for the network controller to overwrite these behaviors and reconfigure the device. 
%%%%%%%%%%%%%%%%%%%%%%%%%%%%%%%%%%%%%%%%%
\subsection{Host Nodes: Software Stack and Middleware Support}
We have previously discussed the design of host nodes based on the principles of software-defined communication. In this approach, each host uses a layered design to separate the concerns necessary for the application-specific transmission and reception of quantum data \cite{HumbleSadlier2014}. We divide this stack  into three distinct layers: software, middleware, and hardware. The software layer consists of the host-specific application as well as the software libraries necessary to access the quantum and classical resource on the node. However, the application need not be aware that communication over a quantum channel occurs, and this design emphasizes the ease of adapting existing applications to use quantum communication. Instead, the middleware layer serves to translate the transmission and reception requests from the software layer into commands that can be parsed by the quantum-enabled hardware. In this regard, the middleware decouples the software from the hardware and removes the requirement that user and node manufacture share a common goal. The price of this design is that the middleware layer must therefore capture knowledge about the range of uses acceptable to hardware \emph{and} conform to an API that is convenient for the software developer. 
\par
The hardware layer exposes access to the low-level physical resources contained within the node. It is responsible for the translation of commands from the middleware layer into hardware specific commands for equipment such as phase modulators and translation stages. An analog to the hardware layer in the traditional computing environment is the device driver, i.e., software that takes commands from the operating system and controls the hardware for a device. For our quantum networking example, this corresponds to the CNIC and QNIC that each connect to respective network links. The hardware interface depends on the physical encoding scheme and transmission media as well as environmental controls and actuators. We have previously tested the layered host design against experimental hardware based on biphoton pair sources and detectors \cite{Pooser2012,Williams2015}. However, we currently focus on the use of numerical simulators as a diagnostic replacement for experimental hardware. The benefit of this approach is two fold. First, we can emulate the functionality of any desired hardware by using numerical simulators of adjustable fidelity. Second, we can use the numerical simulators to efficiently test the behavior of the quantum network as the number of nodes and switches increases. This is both more efficient with respect to labor and materials and more effective in prototyping next generation network applications.
\par
Our numerical model of the hardware layer at each node consists of virtual components that emulate the functionality and properties of expected hardware components. Because we can construct these models to offer the same functionality as the actual hardware layer, we are able to explore the interaction between all three layers prior to experimental testing. In this regard, the middleware is agnostic behavior to the presence of the actual hardware or its numerical model. 
%%%%%%%%%%%%%%%%%%%%%%%%%%%%%%%%%%%
\subsubsection{Switches: Software Stack and Hardware Model}
\begin{figure}[t]
\centering
\includegraphics[width=0.8\columnwidth]{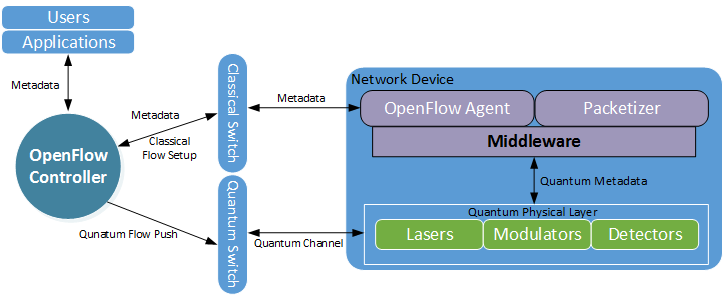}\\
\caption{The interactions between the OpenFlow controller and the classical and quantum switches and the host nodes that support quantum networking.}
\label{fig:qnode}
\end{figure}

The fundamental basis for programmable networks is extracting the control layer for the network away from the data plane that defines how each node operators. The resulting centralized control of the network allows for more powerful management of the networking nodes. In our network design, we use the Ryu OpenFlow controller to define the control layer and to interface with the classical and quantum switch.  As shown in Fig.~\ref{fig:qnode}, the communication between the controller and switches is handled by the OpenFlow protocol, which enables the controller to poll for attributes of the switch. Each switch monitors a series of flow tables that are used to inspect and act upon incoming packets. For the classical switch, these packets inform address resolution and service decisions, while the quantum switch uses the flow tables to parse messages forwarded by the classical network. This approach centralizes the control of both switches within the Ryu controller and permits the controller to coherently manage how the classical and quantum networks interacts. 
%%%%%%%%%%%%%%%%%%%%%%%%%%%%%%%%%%%
%%%%%%%%%%%%%%%%%%%%%%%%%%%%%%%%%%%
As data flows in the network, the switches use their flow tables to make forwarding decisions as packets arrive. The flow tables are built upon flow entries consisting of match fields, counters, and a set of instructions to apply to matching packets. If a matching entry is found, the instructions associated with the specific flow entry are executed at the switch. These instructions include packet forwarding and packet modification. Thus, each flow entry has an action associated with it. The action can forward the packet to a given port, encapsulate and forward the packet to the controller, forward the packet to the next flow table in the pipeline, or drop the packet. 
\par
Quantum metadata, inserted into classical packets, defines the quantum transmission paths. Using software defined networking within our network infrastructure allows us to insert the quantum metadata within network flows observable by the switch. The switch can then alert the controller when quantum metadata is detected and the controller can forward this information to a quantum switch, making it aware of what paths to open for quantum transmission. 
\par
The  design of the quantum switch, in order to communicate with the controller, is built using three layers that are again based on separation of concerns. A classical networking layer within the quantum switch is responsible for interacting with the SDN controller and other classical switches or nodes located on the network. We represent the classical networking layer using a OpenFlow-compatible switch called OpenvSwitch. OpenvSwitch communicates with the OpenFlow controller and also interfaces with the switch middleware layer. The purpose of the middleware is to translate classical actions produced by flow entries within OpenvSwitch to hardware configurations of the quantum switch. The hardware layer represents the quantum optical hardware necessary to route quantum states encoded in single and bi-photon transmission. Because of the no-cloning principles, the switch can not measure and resend the incoming state. Instead, the coherence within the quantum state must be preserved while passing through the switch. A direct method of ensuring coherent in optical states is to employ linear optical elements for the switch physical layer. We impose that requirement on our quantum switches, however, other approaches based on non-linear optical phenomena are also possible.
%%%%%%%%%%%%%%%%%%%%%%%%%%%
\subsubsection{Links: Quantum Optical Characteristics}
The requirements of a quantum optical network link differs from the modern design. The current classical protocol of multiplexed distribution and replication utilizing repeaters does not support transmission of photonic quantum states. A state's fragile coherences and inability to be cloned challenges its inclusion in these networks. The quantum optical hardware must be designed with coherence preservation in mind. For the polarization-entangled quantum states that typify many state of the art experiments, the links within the quantum networking layer must preserve the polarization coherence within a potentially separated photon pair. Our design of quantum physical layer imposes requirements on the use of low-loss network components that preserve this type of polarization coherence. However, in general, there are tradeoffs in the choice of hardware components matching this design. For instance, polarization maintaining fiber is a convenient component in preserving coherence, but is substantially more lossy that standard optical fiber which scrambles the state coherence. Successfully demonstrating a multi-node quantum network necessarily requires some trade-offs between data and error rates. Of course, this is a challenge that classical networks share, and some methods applicable in the classical space are also applicable in the quantum space. For instance, both quantum and classical error correction are useful for mitigating against the decoherence and loss found in conventional components\cite{Sadlier2016}.
\par
We design our quantum networking links using polarization maintaining optical fiber, coherence preserving switches, and precise temporal alignment. Optical fiber links between any two network components are assumed to consist of custom birefringence compensated optical fibers. This supports the coherence of the state by symmetrizing the travel time of photons with different polarizations while maintaining these polarizations. Novel optical switches are utilized that preserve the quantum state up to a known rotation $X,Y,$ or $Z$, are low loss, non-interferometric, and have potential to switch photon paths in less than 100 ns. One mechanism enabling superdense coding is Bell state analysis utilizing Hong-Ou-Mandel interference \cite{hong1987measurement} which requires picosecond timing precision between the members of a photon pair during a Bell state measurement. While transport over large distances challenges this requirement due to difficulties with path length temperature and infrastructure dependencies, for local networks in controlled environments simple calibrations are sufficient.
%%%%%%%%%%%%%%%%%%%%%%%%%%%%%%%%%%%
\section{Quantum Network Simulation}
We use simultaneous simulation of both the classical and quantum network traffic to predict the behavior of a multi-node quantum network. Each network node executes the layered stack described previously using a model for the hardware layer. In this section, we discuss how the simulation environment is constructed using a combination of software tools.
\subsection{Classical Network Simulation}
Mininet is a powerful network simulator that is especially useful for experimenting with Software Defined Network applications. To simply explain, you pass it a file specifying the topology you want to simulate and it establishes the environment using the Linux namespace software. Our topology consists of two virtual switches, three host nodes and an OpenFlow controller, as shown in Fig.~\ref{fig:quantumnetworkdesign}. The communication between the switches and the controller is significant, since this is where how the OpenFlow protocol enforces the network management plane to be extracted from the switches and centralized at the controller. The controller manages both switches and provides there switching logic by accessing their flow tables.
\par
Our switches are built using the OpenvSwitch software and are managed through the OpenFlow protocol by using the Ryu controller. We are bound to the use of version 1.4 of the OpenFlow protocol in order to access optical attributes of OpenFlow. Our controller and switches run on OpenFlow 1.4 version. The program at the switch is a simple switch design that learns MAC addresses and forwards traffic to the proper ports. This program is enabled to run on the switch by directing the switch to communicate with the controller, while the controller gives the switch its forwarding logic. As network states change the switch asks the controller for instructions and the switch updates its flow tables.
\par
\begin{figure}[t]
\centering
\includegraphics[width=1.0\columnwidth]{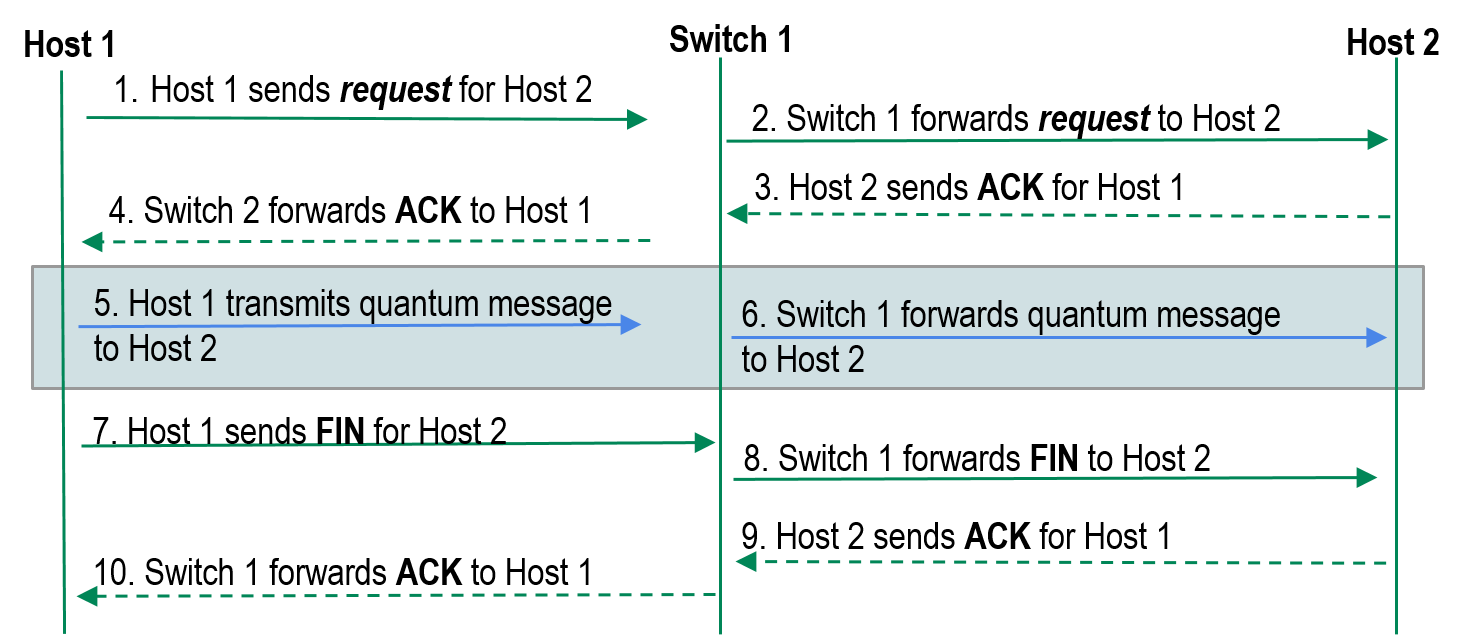}\\
\caption{Interactions between hosts and switch on both quantum and classical channels.}
\label{fig:handshaking}
\end{figure}

In order to confirm that quantum metadata is transferred between nodes we use a classical TCP handshake to set up the connection as shown in Fig.~\ref{fig:handshaking}. Following acknowledgments between hosts, we transfer a quantum message and close the connection. This protocols treats the existence of the quantum message as the metadata communicated by a classical packet and subsequently recognized by the switch flow tables. The switch then sends these specific packets to the controller and the controller relays the proper information to the quantum switch.
%%%%%%%%%%%%%%%%%%%%%%%%%%%%%%%%%%%
%%%%%%%%%%%%%%%%%%%%%%%%%%%%%%%%%%%
%
\subsection{Quantum Network Simulation}
Simulation of the quantum network traffic requires a centralized manager that can monitor the global state of the quantum network physical layer. This is the primary difficulty for quantum simulation since it requires active monitoring of the classical network traffic to inform how each node and link should behave during the simulation. For example, when a bipartite entangled state is distributed across two nodes, its joint state will depend on the local actions taken by each host application. Therefore, the classically-defined instructions issued at each node must be caught by the global simulation manager. Our approach to state simulation is based on centralizing the storage and simulation of quantum systems that exist on the quantum network. We also only monitor the actions that nodes submit to the hardware layer, which has been virtualized in our design for the express purpose of simulation. We use a server-client architecture that permits node to submit operations and receive data to and from a simulation server. We refer to the simulation server as  the \textit{dispatcher} since it controls and dispatches all interactions with the quantum network layer.
\par
By using simulation, the client middleware is able to connect directly to the dispatcher. This places the responsibility of emulating the physical link in the dispatcher, which agrees with our design goal of centralizing the computation of quantum effects. Our middleware implementation uses a simulation specific module that provides a means to connect to the server and supplies additional feedback. An example of this feedback may be the state that the client's quantum detector is in, which will define the classical control signals taht the client receives following a detection event.
\par
The dispatcher has a holistic view of the quantum network. This is beneficial for modeling channel noise that is time dependent and requires accounting for previous transmissions on a particular quantum channel (memory effects). Centralization also eliminates the need of artificial synchronization between clients. The dispatcher stores the entire state of the quantum network and its a history with predefined depth. The result is that determining the history of messages for a channel is trivial based on history lookup.
\par
The numerical simulation of quantum states is separated from the modeling of noise. The dispatcher only determines the operations for the quantum simulation while a separate numerical simulator is used to evaluate these models. Currently, we use a numerical simulation manager \textit{Sabot} to launch and catch results from quantum circuit simulations. Our quantum circuit simulator is based on the {CHP} simulator and its stabilizer formalism \cite{Aaronson2004}. Although CHP is not capable of universal quantum simulation, it does provide an efficient method for simulating the stabilizer circuits that frequently arise in quantum communication protocols. In addition, the Sabot simulation manager provides methods for easily extending these circuit model to other forms of simulation. Sabot also provides several convenience features to improve access and an increase in functionality. This includes a dispatcher facing server, a quantum circuit description interpreter, and a pseudorandom number generator. The Sabot server provides an {API} by which the dispatcher can access internal methods for the creation and modification of quantum states. This supports interactive simulations that can be updated based on changes in network states.
\par
An interpreter function is used to generate a circuit description written in quantum assembly based on QASM into a sequential list of quantum operators. The dispatcher generates the circuit description for simulation after evaluating the appropriate noise model given relevant parameters about the classical and quantum state of the network. This design decouples the specification of a noise model from the numerical simulation and has the effect of increasing the flexibility of the system. 
%%%%%%%%%%%%%%%%%%%%%%%%%%%%%%%%%%%%
\section{Simulating Quantum Network Behavior}
We use the model for a quantum network and the network simulator described above to characterize the behavior of a superdense coding application. The purpose of this demonstration is to presents the integrated function of the individual components. Recall that in superdense coding, Alice and Bob initially share a pair of entangled qubits. Alice has a 2 bit message $b_1b_0$ she wishes to communicate to Bob. She encodes her message by applying one of the four unitary operators $\mathcal{O} \in \{I, X, Z, XZ\}$, which uniquely maps the original state within the complete set of Bell states.
%%%%%%%%%%%%%%%%%%%%%%%%%%%%%%%%%%%%
Alice and Bob establish the specific encoding scheme before transmission. An example of this encoding scheme is presented in table~\ref{tab:encoding}. Alice applies the operator $\mathcal{O}(b_1 b_0)$ to her qubit and transmits it to Bob. Bob must receive both entangled qubits and perform a joint measurement that discriminates between the four Bell states. Based upon the outcome of the measurement, Bob decodes the original message from Alice.
%%%%%%%%%%%%%%%%%%%%%%%%%%%%%%%%%%%
\begin{table}[H]
\centering
\begin{tabular}{| c | c | c | }
\hline $b_1b_0$ & $\mathcal{O}$ & $\ket{\psi_{A,B}}$ \\
\hline 00 & I & $\ket{\Phi^+}$ \\
\hline 01 & X & $\ket{\Psi^+}$ \\
\hline 10 & Z & $\ket{\Phi^-}$ \\
\hline 11 & XZ & $\ket{\Psi^-}$ \\
\hline
\end{tabular}
\caption{An encoding between classical bits and Bell states.}
\label{tab:encoding}
\end{table}
%%%%%%%%%%%%%%%%%%%%%%%%%%%%%%%%%%%
%
\par
In simulating this protocol, we first construct a network topology in which Alice and Bob are connected via a single classical switch. We assume Alice is transmitting a string of many bit pairs. She and Bob then use the classical handshaking protocol described above to frame the transmission of quantum states. We use a version of the handshaking protocol in which Alice and Bob repeat the protocol for each transmitted Bell state. However, Alice and Bob need not know that their communications are utilizing superdense coding or the characteristics of the quantum channel. 
\par
\begin{figure}[t]
\centering
{%%%
\setlength{\fboxsep}{0pt}
\setlength{\fboxrule}{1.5pt}
\color{figbdr}
\fbox{\includegraphics[width=1.0\columnwidth]{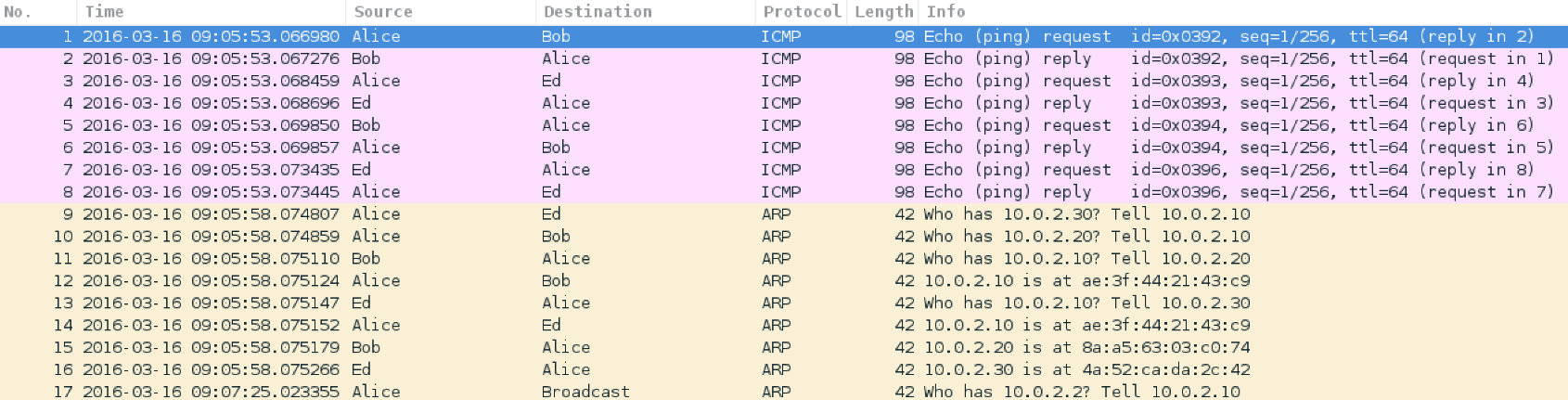}} \\
}%%%
\caption{The traffic between Alice and Bob on the classical network as viewed in Wireshark.}
\label{fig:wireshark}
\end{figure}
\begin{figure}[h]
\centering
\includegraphics[width=1.0\columnwidth]{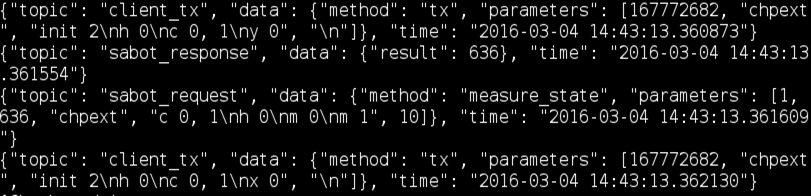}\\
\caption{Events internal to the quantum simulator can be viewed with timestamps to correlate with events on the classical network.}
\label{fig:dspy}
\end{figure}

During the network simulation, mininet tracks the state of the classical packets and simulates the arrival at the hosts and switch. We use wireshark, a popular and free network packet analyzer, to monitor the handshaking traffic between Alice and Bob occurring on the classical network within mininet. Example output from Wireshark is shown in figure~\ref{fig:wireshark}, for which messaging between Alice and Bob is apparent. Also shown are network messages between the host nodes and \emph{ED}, the dispatcher host. This traffic represent the quantum network simulation traffic.
%%%%%%%%%%%%%%%%%%%%%%%%%%%%%%%%%%%
%%%%%%%%%%%%%%%%%%%%%%%%%%%%%%%%%%%
We developed an analogous diagnostic tool for monitoring the state of the quantum network simulated within the dispatcher. An example of this textual output is in Fig.~\ref{fig:dspy}. This diagnostic tool captures events from the quantum simulator dispatcher and transmit them to a remote node for processing within Sabot. The message data s presents the simulation metadata as well as quantum state or measurement results that are communicated between the network hosts and dispatcher. For example, note that timestamps used to account for time correlations between events on the quantum network and the classical network. This is particularly useful for network hosts attempting non-local protocols based on time-arrival of the transmissions.
%%%%%%%%%%%%%%%%%%%%%%%%%%%%%%%%%%%
%%%%%%%%%%%%%%%%%%%%%%%%%%%%%%%%%%%
\par
After transmission is complete, Bob has the original message Alice intended to communicate. If the quantum channel simulated were to include noise in the form of decoherence or losses, then Bob would receive Alice's original message with some errors. These errors could be mitigated with quantum error correction or in the case of SDC, classical forward error correction could be used \cite{Sadlier2016}. The latter is beneficial because it does not require larger quantum states, which is challenging to achieve with current technology and is computationally intensive to simulate. The action of classical forward error correction could take place in the data processing component of the software stack allowing for changes to be made to the error correction scheme without affecting the requirements of the components associated with quantum transmission and reception.
\par
While this demonstration has emphasized the interactions between two hosts, Alice and Bob, the classical and quantum network simulators can be easily extended to add additional nodes to the network. It is our hope that these tools can observe the emergent behavior of the network that will allow the engineering of robust communication protocols that mitigate the effects of, for example, transmission collisions, heavy network traffic, and noise communication channels.
\section{Conclusions}
We have presented the first design of a programmable quantum network using the principles of software-defined networking. Our approach has realized implementations for hosts and switches that support quantum communication including network interfaces for both the classical and quantum networking layers. We have introduced the OpenFlow controller to manage the interaction between the classical and quantum networking layers by setting the different behaviors of these heterogeneous switches. In addition, we have developed numerical simulators capable of modeling the complete quantum network including its classical metadata and quantum state. We have leveraged the mininet simulation environment for realizing the classical network communication while we have implemented a numerical simulator to track the state of the quantum network layer. Finally, we have applied this simulator to the case of superdense coding using entangled photons and we have discussed how it can be used to test novel ideas for error corrected detection.
\par
Using programmable network principles to manage the behavior of the quantum network offers additional opportunities for managing its heterogeneous nodes. While our design has focused on the development of nodes supporting quantum optical hardware, these designs can be easily modified to accommodate other hardware layers. This is because details regarding the hardware and its physics are abstracted by the separation of concerns and especially the middleware layer, which exposes logical features that can be realized. For quantum optical hardware, transmitting and receiving are natural logical behaviors that may be exposed to the controller. By contrast, trapped ion hardware offers capabilities for memory and logical processing within the node. The controller can manage these devices differently according to the logical features as opposed to the hardware features. This permits more robust design of the network especially with respect to modification or upgrades of the node hardware.
%%%%%%%%%%%%%%%%%%%%%%%%%%
\begin{acknowledgements}
This work was supported by a research collaboration with Oak Ridge National Laboratory and US Army Research Laboratory. VRD expresses his gratitude to the OSD Applied Research for the Advancement of S\&T Priorities (ARAP) Program for its partial financial support of this work.
\end{acknowledgements}

\bibliographystyle{spiebib}
\bibliography{spieqic2016_manuscript}

\begin{thebibliography}{10}

\bibitem{VanMeter2014}
{V}an {M}eter, R.,  [{\em Quantum Networking}{\nolinebreak\hspace{0.1em}]},
  John Wiley and Sons (2014).

\bibitem{Britt2015}
Britt, K.~A. and Humble, T.~S., ``High-performance computing with quantum
  processing units,'' {\em preprint} (arXiv:1511.04386) (2015).

\bibitem{Gisin2002}
Gisin, N., Ribordy, G., Tittel, W., and Zbinden, H., ``Quantum cryptography,''
  {\em Rev. Mod. Phys.}~{\bf 74},  145--195 (2002).

\bibitem{Broadbent2009}
Broadbent, A., Fitzsimons, J., and Kashefi, E., ``Universal blind quantum
  computation,'' in [{\em Proceedings of the 50th Annual {IEEE} Symposium on
  Foundations of Computer Science (FOCS 2009)}{\nolinebreak\hspace{0.1em}]},
  517–--526 (2009).

\bibitem{Humble2013}
Humble, T.~S., ``Quantum security for the physical layer,'' {\em Communications
  Magazine, IEEE}~{\bf 51}(8),  56--62 (2013).

\bibitem{Williams2015}
Williams, B.~P., Britt, K.~A., and Humble, T.~S., ``A tamper-indicating quantum
  seal,'' {\em Phys. Rev. Appl}~{\bf 5} (2015).

\bibitem{Campbell1999}
Campbell, A.~T., De~Meer, H.~G., Kounavis, M.~E., Miki, K., Vicente, J.~B., and
  Villela, D., ``A survey of programmable networks,'' {\em ACM SIGCOMM Computer
  Communication Review}~{\bf 29}(2),  7--23 (1999).

\bibitem{Hu}
Hu, F., Hao, Q., and Bao, K., ``A survey on software-defined network and
  openflow: From concept to implementation,'' {\em {IEEE} Communication Surveys
  and Tutorials}~{\bf 16}(4) (2014).

\bibitem{HumbleSadlier2014}
Humble, T.~S. and Sadlier, R.~J., ``Software-defined quantum communication
  systems,'' {\em Optical Engineering}~{\bf 53}(8),  086103 (2014).

\bibitem{Dasari2015}
Dasari, V.~R. and Humble, T.~S., ``Openflow arbitrated programmable network
  channels for managing quantum metadata,'' {\em preprint} (arXiv:1512.08545)
  (2015).

\bibitem{Pooser2012}
Pooser, R.~C., Earl, D.~D., Evans, P.~G., Williams, B., Schaake, J., and
  Humble, T.~S., ``Fpga-based gating and logic for multichannel single photon
  counting,'' {\em Journal of Modern Optics}~{\bf 59} (2012).

\bibitem{Sadlier2016}
Sadlier, R.~J. and Humble, T.~S., ``Superdense coding interleaved with forward
  error correction,'' {\em preprint} (arXiv:1601.06321) (2016).

\bibitem{hong1987measurement}
Hong, C., Ou, Z., and Mandel, L., ``Measurement of subpicosecond time intervals
  between two photons by interference,'' {\em Physical Review Letters}~{\bf
  59}(18),  2044 (1987).

\bibitem{Aaronson2004}
Aaronson, S. and Gottesman, D., ``Improved simulation of stabilizer circuits,''
  {\em Phys. Rev. A}~{\bf 70},  052328 (Nov 2004).

\end{thebibliography}

\end{document}